\documentclass[12pt]{iopart}

\usepackage{iopams}  
\usepackage{graphicx}  
\usepackage{color}
\usepackage{cite}
\begin{document}

\title[]{High-order harmonic generation driven by inhomogeneous plasmonics fields spatially bounded: influence on the cut-off law}

\author{E Neyra$^1$, F Videla$^1$, M F Ciappina$^2$, J A P\'erez-Hern\'andez$^3$, L Roso$^3$, M Lewenstein$^{4,5}$ and G A Torchia$^1$ }

\address{$^1$Centro de Investigaciones \'Opticas (CIOp) CONICET La Plata-CICBA, Camino Centenario y 506, M.B. Gonnet, CP 1897, Provincia de Buenos Aires, Argentina}
\address{$^2$Institute of Physics of the ASCR, ELI-Beamlines, Na Slovance 2, 182 21 Prague, Czech Republic}
\address{$^3$Centro de L\'aseres Pulsados (CLPU), Parque Cient\'{\i}fico, E-37008 Villamayor, Salamanca, Spain } 
\address{$^4$ICFO - Institut de Ciencies Fotoniques, The Barcelona Institute of Science and Technology, Av. Carl Friedrich Gauss 3, 08860 Castelldefels (Barcelona), Spain }
\address{$^5$ICREA, Pg. Llu\'is Companys 23, 08010 Barcelona, Spain }

\ead{gustavot@ciop.unlp.edu.ar}

\begin{abstract}

We study high-order harmonic generation (HHG) in model atoms driven by plasmonic-enhanced fields. These fields result from the illumination of plasmonic nanostructures by few-cycle laser pulses. We demonstrate that the spatial inhomogeneous character of the laser electric field, in a form of Gaussian-shaped functions, leads to an unexpected relationship between the HHG cutoff and the laser wavelength. Precise description of the spatial form of the plasmonic-enhanced field allows us to predict this relationship. We combine the numerical solutions of the time-dependent Schr\"odinger equation (TDSE) with the plasmonic-enhanced electric fields obtained from 3D finite element simulations. We additionally employ classical simulations to supplement the TDSE outcomes and characterize the extended HHG spectra by means of their associated electron trajectories. A proper definition of the spatially inhomogeneous laser electric field is instrumental to accurately describe the underlying physics of HHG driven by plasmonic-enhanced fields. This characterization opens new perspectives for HHG control with various experimental nano-setups.
 
\end{abstract}


\vspace{2pc}


\maketitle

%

\section{Introduction}

Since the seminal experiment performed at the group of A. L'Huillier~\cite{lhuiller}, the high-order harmonic generation (HHG) process constantly increases its relevance and also provides important tools to conduct fundamental experiments in atomic and molecular systems. The nonlinear interaction of ultra-short intense laser pulses with atoms or molecules produces a plethora of phenomena~\cite{protopapas,brabec,krausz}.  Amongst them, the HHG, i.e.~the generation of coherent radiation in the range of extreme ultraviolet (XUV) to Soft-X-Ray spectral range, is one of the most widely experimentally used and theoretically studied. The HHG phenomenon can be condensed using the 3-step model~\cite{corku93A,lewenstein94A,schaf93A}, namely, (i) in the first step an electronic wavepacket is sent to the continuum by tunnel ionization through the potential barrier created as a consequence of the non-perturbative interaction between the atom and the laser electric field; (ii) after that, in the second step, the laser-ionized electronic wavepacket propagates in the continuum and is driven back to the vicinity of the parent ion when the laser electric field reverses its direction; (iii) finally, in the third stage of the sequence, this electronic wavepacket has certain probability to recombine, producing an ultrashort -of the order of hundreds of attoseconds, burst of coherent harmonic radiation after the electron-ion recombination.
Both classical and quantum approaches predict that the maximum harmonic order (the so-called HHG cutoff) can be estimated by: \begin{equation}
\label{cutoff}
n_{c}=(3.17 U_p+I_p)/\omega
\end{equation}
where $I_p$ is the ionization potential of the atom or molecule under consideration and $U_p$ is the ponderomotive energy, given by $U_p=E_0^2/4\omega^2$, $E_0$  being the laser electric field peak amplitude and $\omega$ the laser central frequency. Note that $U_p\propto I\lambda^2$ where $I$ and $\lambda$ are the laser intensity and wavelength, respectively. According to Eq.~(\ref{cutoff}) there are two possible routes to extend the HHG cutoff $n_c$ (i.e.~to produce photons with much higher energy), namely, (a) to increase the laser wavelength $\lambda$~\cite{spielmann,tenio} or (b) to use more intense lasers. It must be mentioned that, however, longer wavelengths implies a sudden decrease in the HHG efficiency, governed by the well-known $\lambda^{-5.5}$ law \cite{tate_07A,frolov_08,perez-hernandez09A}. On the other hand, the utilization of more powerful lasers has a fundamental drawback: very rapidly we reach the saturation regime and then the atom or molecule results completely depleted within the leading edge of the laser pulse. This clearly results in a dramatic reduction in the HHG yield (or, in the most extreme case, in a total absence of emitted harmonic radiation)~\cite{strelkov,jose_oe,moreno1995}.

HHG obtained by means of the interaction of ultra-short laser pulses with noble gases has become in the last decade a key phenomenon to produce trains or isolated attosecond pulses~\cite{farkas92,hentschel01}. Many efforts are conducted by the scientific community in order to improve and optimize the HHG process~\cite{perez-hernandez09A,carrera,chipperfield09,serrat,enriquelp,enriqueepjd}; related to this aim, a recent milestone was achieved: by taking advantage of the laser field enhancement obtained by localized plasmons polaritons (LPP), it was shown that is possible to generate HHG with larger cutoffs. The LPP appears when a metal nano-structure is illuminated by a short laser pulse~\cite{kim08} and the resulting plasmonic-enhanced field presents two main characteristics, namely (i) it is an amplified replica of the incoming, typically low intensity -$10^{11}\sim10^{12}$ W/cm$^{2}$, driving laser pulse and (ii) it develops spatial variations at a nanometric scale~(for a recent review see e.g.~\cite{ropp}). Besides the initial enthusiasm, the Kim's experiment is not free of controversies (see e.g.~\cite{sivise1,kime2,sivis2013}) and nowadays there exists a consensus that their setup is not the most viable route to generate HHG with low intensity laser sources. The dispute was, however, partially settled in~\cite{han2016}, where low order harmonics were generated with a similar setup, but using a solid piece of material as a driven medium instead of a rare gas. From a theoretical viewpoint, the spatially nonhomogeneous character of the plasmonic-enhanced fields has a tremendous influence in the underlying physics of HHG~\cite{prl,cpc}. The effect of the plasmonic-enhanced electric fields can be accounted for by including a function that models their spatial variation. It was demonstrated that a clear cutoff extension appears and it is mainly due to the spatial inhomogeneous character of the driving field~\cite{husakou11, yavuz12,marcelo12,marcelooe}. In particular, in Ref.~\cite{marcelooe} it was pointed out that the precise description of the spatial form of the plasmonic-enhanced fields is essential for an accurate prediction of the cutoff extension.


In this contribution we study the fundamental physics of HHG in a model atom driven by a spatially inhomogeneous plasmonic-enhanced electric field. As was stated, this field results from the interaction between an ultra-short laser pulse and a metal nanostructure. In particular we explore the HHG process by using a realistic fitting of the electric near-field produced by a gold bow-tie shaped nanostructure. The results obtained in this paper are called to be very useful by considering they can clarify some aspects that have not been assessed until now. For instance, the utilization of a Gaussian-shaped and bounded functional form for the spatial variations of the plasmonic-enhanced electric field presents many advantages, that we describe throughout this article. Our quantum mechanical calculations, obtained by solving the one-dimensional time-dependent Schr\"odinger equation (1D-TDSE), in an argon model atom, are validated by using a pure classical analysis. In addition, further insights about the electron trajectories driven by these spatially bounded inhomogeneous electric fields, joint with the reasons of the HHG cutoff extension, can be extracted from this approach as well.


Our findings can be summarized as follows: we show that (i) the HHG cutoff energy scales with the laser wavelength at a power larger than 3. We also give an explanation of the reasons of this new dependence. We should emphasise, however, that this conclusion is obtained in a limited excitation spectral range and for a particular bow-tie shaped nanostructure; (ii) there exist upper bounds for the HHG cutoff energy and the spatial range of electron's trajectories. Interestingly, these limits are independent of the excitation wavelength.  In contrast, they are strongly dependent on the plasmonic-enhanced electric field spatial distribution; (iii) the field distribution can be designed tailoring the nanostructure geometry. As a consequence, it appears feasibe to manipulate the HHG features engineering it. This finding can be extended to any plasmonic-enhanced spatially inhomogeneous field, independently of its origin. Moreover, we stress out that the description of the physics involved in our simulations is performed in a simple and comprehensive manner. In addition, both the quantum and classical results agree very well for the complete set of results obtained.

The rest of the article is organized as follows. In the next section, Sec.~2, we discuss the underlying physics of the plasmonic field enhancement and how to compute the spatial shape of the plasmonic-enhanced spatially inhomogeneous field. Our studies include, as much as possible, realistic parameters. In Sec.~3 we briefly describe our quantum mechanical model. We use this approach in Sec.~4 to compute HHG for a large set of parameters. We also discuss here the scaling laws that govern the HHG cutoff in different laser wavelength ranges and what they imply. Finally, in Sec.~5, we summarize the main findings, present our conclusions and give a short outlook. Atomic units are used, unless otherwise stated.

\section{Plasmonic-enhanced fields properties}

\begin{figure}[h]
\centering
\includegraphics[width=0.7\textwidth]{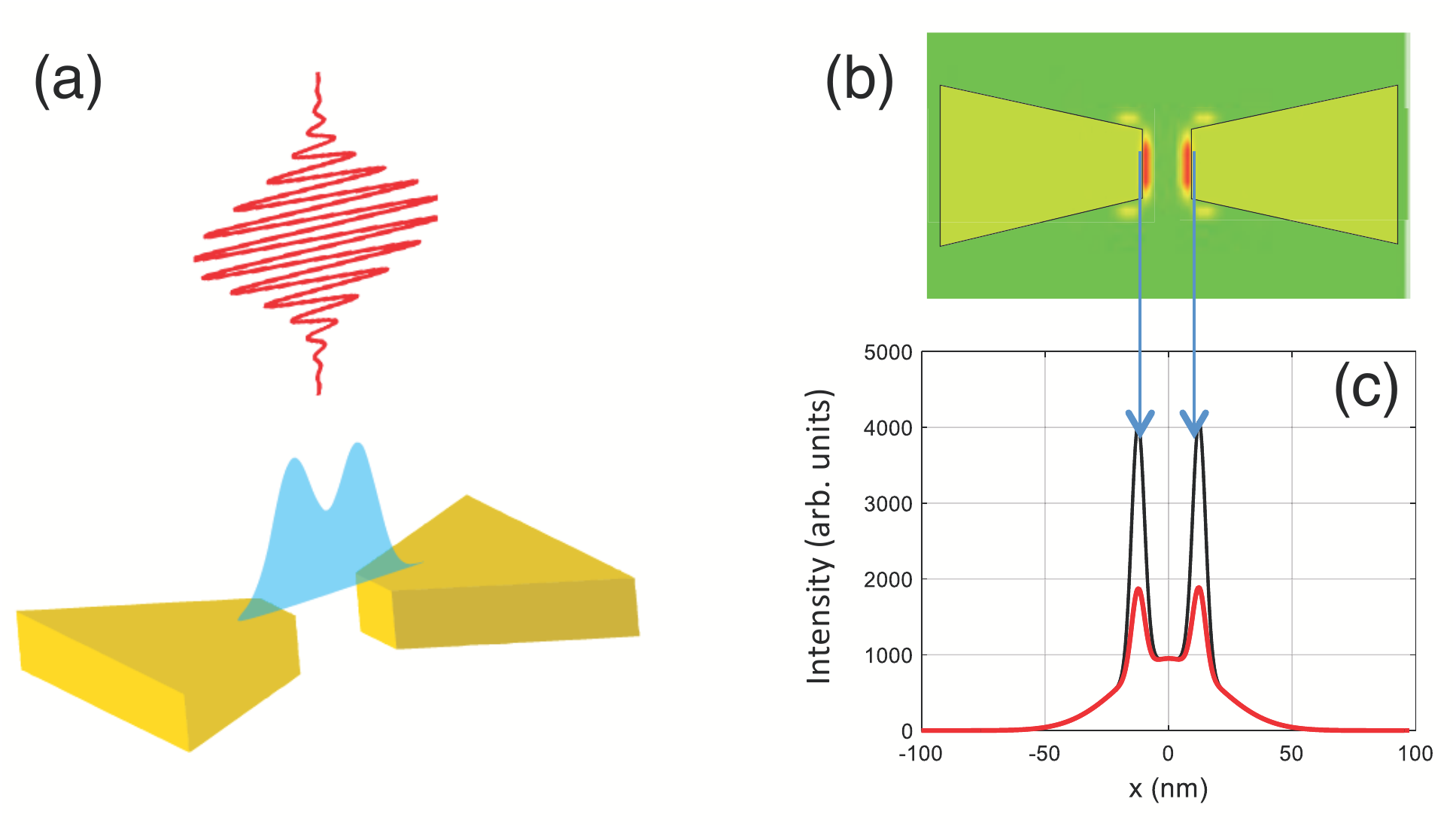}
\caption{(a) Schematic representation of the laser-nanostructure interaction. It is shown (in blue) a pictorial representation of the plasmonic-enhanced field. (b) and (c) represent the simulated electric field produced around the metallic bow-tie nanostructure. In (c) a numerical fit (red line) from the calculated field (black line) obtained from the FDTD simulations is shown. The HHG is generated when the plasmonic-enhanced field interacts with the gas atoms (not shown) that surround the nanostructure (see the text for more details).}
\label{fig1} 
\end{figure}

As it is well described in the paper of Kim {\em et al}~\cite{kim08}, the electric field produced by a bow-tie shaped metal nanostructure, follows the pattern and intensity distribution as sketched in the Fig.~\ref{fig1}(a). As it can be seen, when a laser light field impinges onto the nanostructure, a bi-dimensional electric near-field is built up, with similar temporal characteristics as the incoming laser pulse, but with an enhanced peak amplitude (around 1-2 orders of magnitude, see e.g.~\cite{kim08}).


Additionally, the local near-field distribution is concentrated around the metallic nanostructure and spatially confined (see Fig.~\ref{fig1}(b) for details). The spatial electric near-field distribution was obtained by using Full Wave, a numerical simulation software based on the finite difference time domain (FDTD) method, under the RSOFT suite~\cite{Rsoft}. The longitudinal dimension of each bow-tie was set to 175 nm and in Fig.~\ref{fig1}(c) we display the electric near-field amplitude distribution in a range around $\pm100$ nm centered at the middle of the gap. The intensity profile shown in Fig.~\ref{fig1}(c) was calculated for a laser wavelength $\lambda=800$ nm.  

For our purposes it is enough to reduce the study by considering the cross section between the bow-tie apexes plotted in Fig.~\ref{fig1}(b). This point of view simplifies the representation of the electric near-field in one spatial dimension as it is detailed in Fig.~\ref{fig1}(c), allowing to obtain the HHG yield by means of dimensionally reduced classical and quantum approaches. We have represented the effective near-field in a spatial range where a considerable volume of gas atoms can be ionized and, as a consequence, the HHG yield is strong enough to be experimentally detected. Additionally it is possible to use the same effective near-field shape for other laser wavelengths, because the numerically calculated field distribution for each wavelength changes only its peak amplitude, leaving its horizontal spatial extension practically unchanged. Finally, note that the plasmonic-enhanced electric near-field oscillates at $\omega$, the frequency of the laser driving field. 

\begin{figure}[h]
\centering
\includegraphics[width=0.7\textwidth]{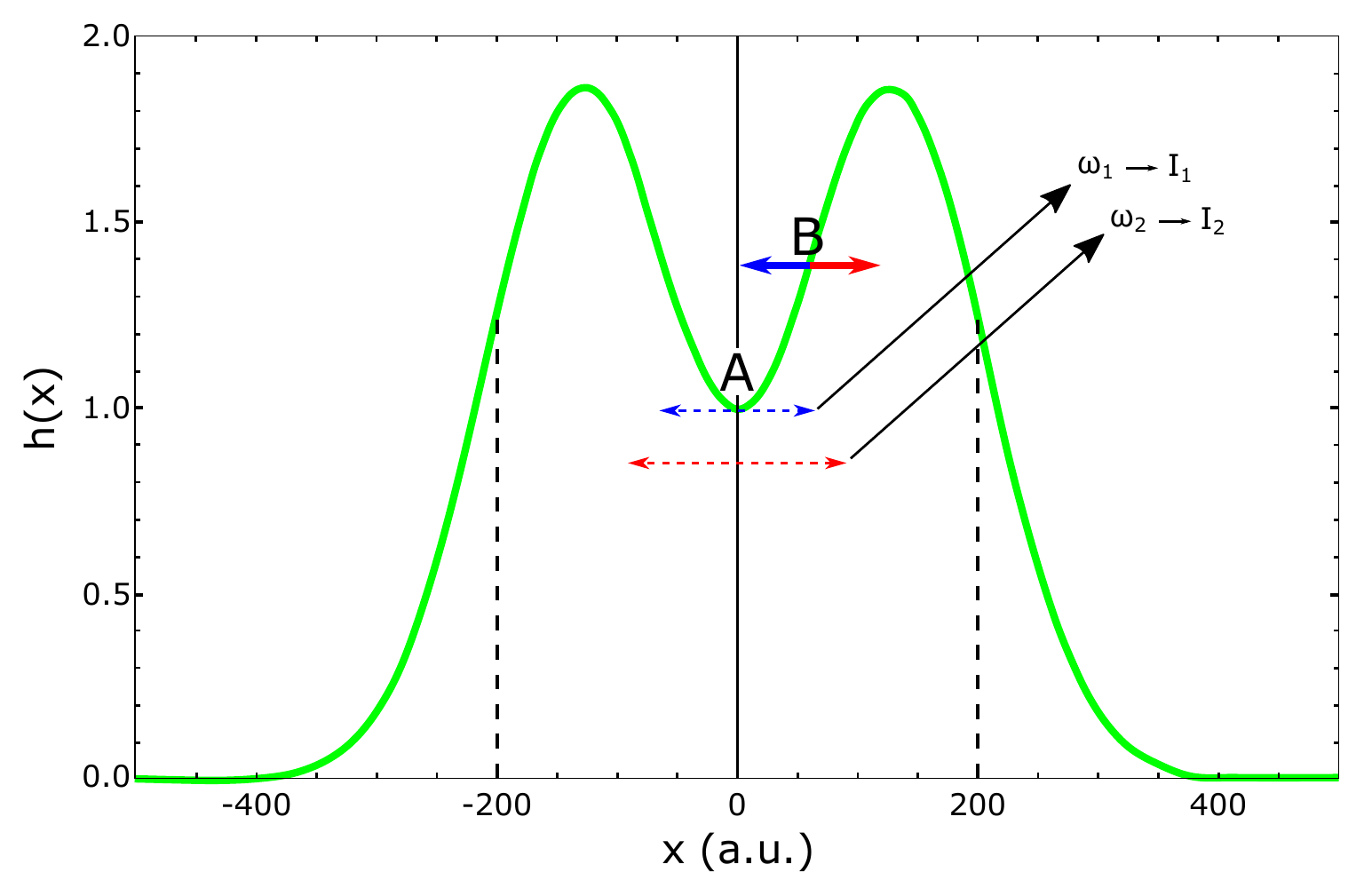}
\caption{Spatial dependence of the plasmonic-enhanced electric near-field. This shape mimics the numerical fit of Fig.~1(c) using two Gaussian functions (see the text for details). We show two zones of interest, namely, (i) the surrounding of the point A, where the field amplitude increases at both sides of it; (ii) a region around the point B, where the field increases/decreases monotonically, depending if the electron moves to the left/right (see the text for more details).}
\label{fig2} 
\end{figure}

Figure~\ref{fig2} shows the one-dimensional dependence of the plasmonic-enhanced electric near-field (note that the field amplitude is normalized at the point A). This curve was constructed by using two Gaussian functions to fit the black curve shown in Fig.~\ref{fig1}(c). As we will show next, an accurate spatial description of the plasmonic driven field is instrumental to understand some of the HHG features.

Point A defines a local minimum of the plasmonic-enhanced electric near-field in the one-dimensional landscape. If we consider that ionization either occurs for both positive or negative electric fields (which correspond in each case to an electron moving to the negative and positive $x$ coordinate, respectively), the electron will always increase its velocity after ionization because at both sides of the point A, the electric field increments its absolute value.  It is important to stress that laser-ionized electrons by means of lower frequencies (larger wavelengths) laser sources travel larger distances in the continuum before recombination (dashed color arrows in Fig.~\ref{fig2}). This indicates they could reach far away regions from their starting initial points and spend larger times in the so-called laser continuum (see e.g.~\cite{marcelooe} for a detailed study about this aspect). As we will see in the next section, these circumstances are the main responsible of the HHG cutoff extension. This is so by considering that, at larger wavelengths, laser-ionized electrons are moving under the influence of an electric field that increases its intensity as a function of the spatial coordinate. Besides, taking into account a Gaussian-shaped profile for the plasmonic-enhanced driving near-field, the field amplification possesses a  `natural' limit, giving thus origin to distinct results.

On the other hand, for the point B, located at  $\sim70$ a.u. in our particular example (see Fig.~\ref{fig2}), the electric near-field amplitude increases (decreases) if we move to the right (left) (solid color arrows in Fig.~\ref{fig2}). HHG driven by plasmonic-enhanced near-fields with a spatial shape restricted to the A-B region was studied in previous works~\cite{husakou11,yavuz12,marcelo12}, where a first order approximation was employed. Below we will show results that confirm and clarify this description in a more detailed form.
 
In principle, one would like to model the plasmonic-enhanced HHG process in a complete way, i.e.~including both single-atom responses and collective effects. This task, however, is not under the reach of nowadays computational resources.
One alternative way to proceed, in order to consider an 'averaged' effect of the plasmonic-enhanced field, would be to use a linear or quadratic model to describe the near-field for each position where the model atom is located. In this way the actual shape of the plasmonic-enhanced field would be entirely `probed'. A procedure following the latter reasoning was used in~\cite{yavuz2013}. One of the main problems of these simple models, however, is that they would break down for larger wavelengths ($\gtrsim 1500$ nm) and higher laser intensities ($\gtrsim 1\times10^{14}$ W/cm$^2$). In fact, laser sources with the latter characteristics are nowadays available and, as a consequence, theoretical models capable to make realistic predictions would be needed.

\section{Theoretical models}

For a linearly polarized laser field, that is the case of our study, the dynamics of an atomic electron is mainly along the polarization direction. As a result, it is indeed a good approximation to employ the time-dependent Schr\"odinger equation in one spatial dimension (1D-TDSE). We can thus write (note that we consider the laser linearly polarized along the $x$-axis) :
\begin{eqnarray}
i\frac{\partial \Psi}{\partial t}&=&\mathcal{H}(t)\Psi(x,t)\\
&=&[-\frac{1}{2}\frac{\partial^2}{\partial x^2}+V_{\mathrm{atom}}(x)+V_{\mathrm{plasm}}(x,t)]\Psi(x,t),
\label{eq:eq2}
\end{eqnarray}
where $V_{\mathrm{atom}}(x)$ is the atomic potential and $V_{\mathrm{plasm}}(x,t)$ represents the potential due to the oscillating plasmonic-enhanced near-field at laser frequency $\omega$, developed by the nanostructure. In here, we use for $V_{\mathrm{atom}}(x)$ the quasi-Coulomb potential:
\begin{equation}
V_{\mathrm{atom}}(x)=\frac{-1}{\sqrt{x^2+\xi^2}},
\label{eq:eq3}
\end{equation}
which was first introduced in~\cite{potential} and has been widely used in 1D studies of laser-matter processes in atoms. The required ionization potential can be defined by varying the parameter $\xi$ in Eq.(~\ref{eq:eq3}). The potential $V_{\mathrm{plasm}}(x,t)$ due to the plasmonic-enhanced electric near-field $E(x,t)$ is given by:
\begin{equation}
V_{\mathrm{plasm}}(x)=-\int E(x,t)dx.
\label{eq:eq4}
\end{equation}
Here the spatial dependency of $E(x,t)$ is strong (see Fig.~\ref{fig2}) and we can incorporate it in $E(x,t)$ as follows: 
\begin{equation}
E(x,t)=E_0f(t)h(x)\sin(\omega t).
\label{eq:eq5}
\end{equation}
In Eq.~(\ref{eq:eq5}), $E_0$, $\omega$ and $f(t)$ are the electric field peak amplitude, the frequency of the coherent electromagnetic radiation and the pulse envelope, respectively. Furthermore, $h(x)$ represents the functional form of the spatial nonhomogeneous part of the plasmonic-enhanced electric near-field. This spatial dependence can be represented as a sum of two gaussian functions, that in turn can be written as a series of the form:
\begin{equation}
h(x)=\sum_{i=0}^n b_ix^i,
\label{eq:eq6}
\end{equation}
where the coefficients $b_i$ are obtained by fitting the real electric field shape (see Fig.~2). This field results from FDTD simulations, considering the real geometry of the bow-tie shaped nanostructure. In order to obtain a good fitting of the function $h(x)$, we have developed the expansion, Eq.~(\ref{eq:eq6}), up to the fortieth order ($n=40$). Throughout this work we use a pulse envelope $f(t)$ of the form:
\begin{equation}
f(t)=\sin^2\left(\frac{\omega t}{2n_p}\right),
\label{eq:eq7}
\end{equation}
where $n_p$ is the total number of cycles, set as $n_p=6$. 
We use $\xi= 1.18$ in Eq.~(\ref{eq:eq3}) such that the binding energy of the ground state of the 1D Hamiltonian coincides with the (negative) ionization potential of argon, i.e.~$I_P =-15.76$ eV ($-0.58$ a.u.). Moreover, we assume that the noble gas atom is in its initial ground state (GS) before $(t=-\infty)$ we turning the laser on. Equation~\ref{eq:eq2} is then numerically solved by using the Crank-Nicolson scheme. In addition, to avoid spurious reflections from the spatial boundaries, at each time step, the electron wavefunction is multiplied by a mask function. In our case, both the nanostructure gap (see e.g.~\cite{marcelo12}) and the spatially bounded character of the plasmonic-enhanced field, limit the electron motion and, in turn, the numerical spatial grid.
For instance, at $I=1\times10^{14}$ W/cm$^{2}$ and the largest wavelength employed, $\lambda=1500$ nm, we have used $x_{lim}\sim\pm 400$ a.u., with a spatial step of 0.04 a.u. (around 20 000 spatial points). The atomic harmonic yield is then computed by Fourier transforming the so-called dipole acceleration $a(t)$ of its active electron. That is,
\begin{equation}
D(\omega)\propto\left|\int_{-\infty}^{\infty}dt e^{-i\omega t}a(t)\right|^2,
\label{eq:eq8}
\end{equation} 
where $a(t)$ is obtained by using the following commutation relationship:
\begin{equation}
a(t) = \frac{d^2 \langle x \rangle}{dt^2} =\langle{\Psi(x,t)}|[\mathcal{H}(t),[\mathcal{H}(t),x]]|{\Psi(x,t)}\rangle.
\label{eq:eq9}
\end{equation}
In here, $\mathcal{H}(t)$ and $\Psi(x,t)$ are the Hamiltonian and the electron wavefunction defined in Eq.~(\ref{eq:eq2}), respectively. The function $D(\omega)$ is also called dipole spectrum, which gives the spectral HHG profile measured in experiments.

\section{Results and discussion}

\begin{figure}[h]
\centering
\includegraphics[width=0.7\textwidth]{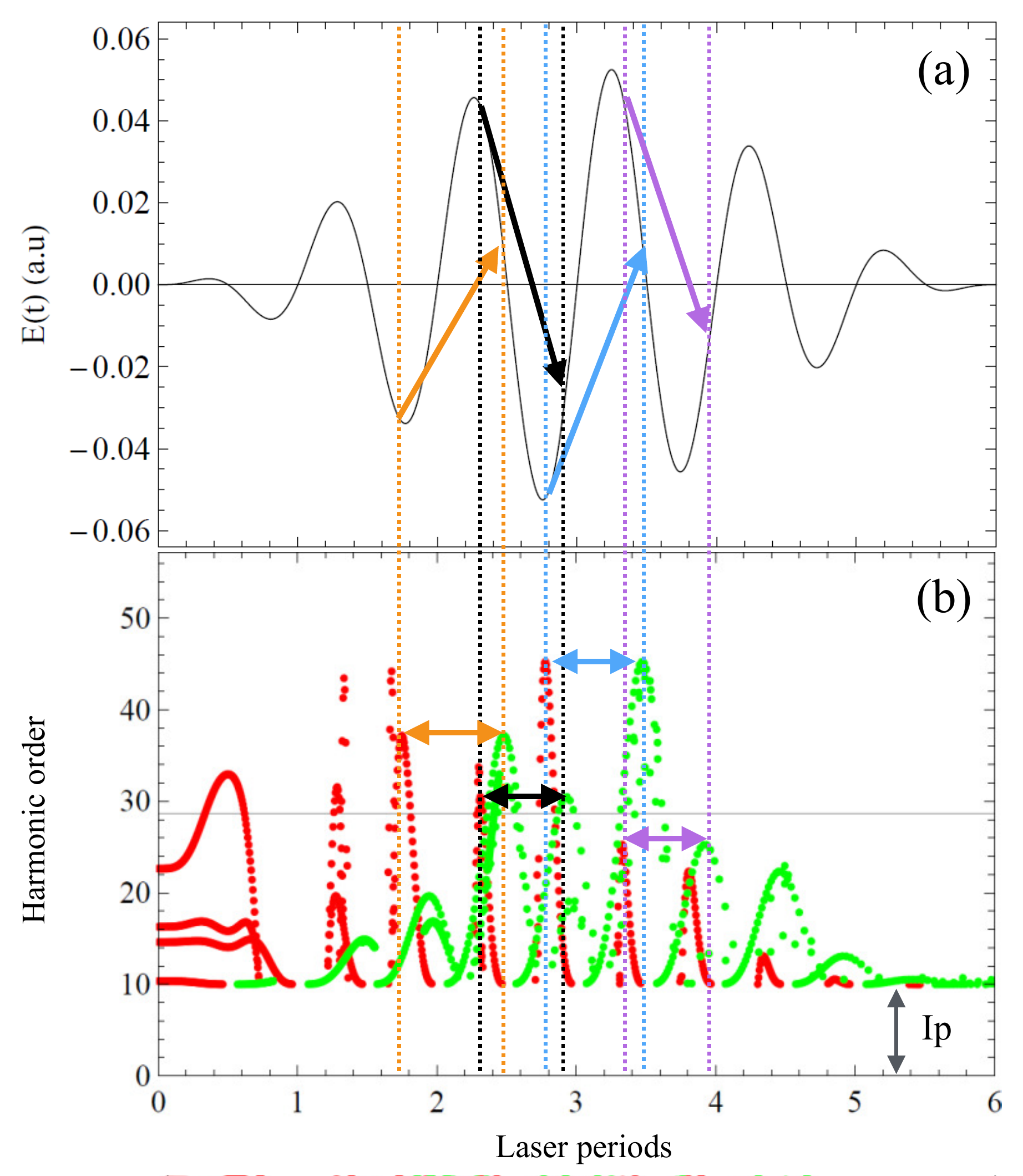}
\caption{(a) Temporal profile of the driving laser field. The laser enhanced intensity and wavelength are $I=1\times10^{14}$ W/cm$^{2}$ and $\lambda=800$ nm, respectively. The laser pulse comprises 6 total cycles and the coloured arrows indicate the times where ionization (tail) and recombination (tip) occur. (b) Classical  electron energies (in terms of the harmonic order) at recombination as a function of the ionization (red dots) and recombination times (green dots) for a model atom located at point B. The vertical dashed coloured lines link up the temporal intervals denoted in panel (b) by the horizontal arrows with the corresponding ionization-recombination instants along the laser pulse in panel (a). The color criterion of the arrows and vertical lines connects the corresponding events in both panels. In addition the $I_P$ of the argon atom, $15.76$ eV (0.58 a.u), is shown.}
\label{fig3} 
\end{figure}

The following calculations are made using a laser wavelength $\lambda=800$ nm and considering an atom located at the point B (see Fig.~\ref{fig2}). In this way we will first support and validate the linear approximation for the plasmonic-enhanced near-field~\cite{marcelo12}. This assumption can be considered valid because the movement of the electron, at this laser wavelength, is developed in a tiny spatial region. We show from these results the importance of the sign of the electron motion under spatial inhomogeneous fields. In Fig.~\ref{fig3} the classical harmonic order vs. the ionization/recombination time is shown (for details see e.g.~\cite{cpc}). The coloured arrows detail, for each part of the pulse (Fig.~\ref{fig3}(a)), the trajectories followed by the electrons before and after ionization. It would be worth mentioning that, considering the electron electric charge, a negative electric field pushes the electron far away meanwhile the positive part of the driving pulse brings the electron back to the vicinity of the parent ion (recombination). As a consequence, it is expected that higher-order harmonics can be generated when the electron is ionized by the negative part of the driving pulse. To clarify this phenomenon, different colour arrows are used in Fig.~\ref{fig3}(b) to describe the above-mentioned process, at different cycles for the driving pulse. In order to obtain a larger cutoff, it is necessary, in all the cases, to take into account both the intensity and direction for each part of the driving pulse as it is detailed in Fig.~\ref{fig3}. We can also observe from Fig.~\ref{fig3} that different energies at the recombination are obtained, depending on the direction of electrons movement (see Fig.~\ref{fig2}).  As an example, this fact can be clearly observed comparing the energy obtained for electrons ionized by the part of the pulse marked by a red arrow. In this case the electrons travel to positive $x$ coordinates, where there is a large electric field amplitude, and the recombination energy is higher when compared with that for electrons ionized by the part of the pulse indicated by a black arrow (corresponding to negative $x$ coordinates, where the electric field amplitudes are smaller). The above description put forward the differences between the underlying physics of the HHG driven by spatial homogeneous and inhomogeneous fields. A similar set of results was reported by Shaaran \textit{et al}~\cite{tahir12} using a quantum orbits analysis.

The next study is performed putting the model atom at $x=0$ and at a laser-enhanced intensity of $1\times10^{14}$ W/cm$^2$ (see the point A in Fig.~\ref{fig2}). For all simulations we have considered a pulse given by Eq.~(\ref{eq:eq5}). In order to analyze the evolution of the HHG cutoff with the laser wavelength $\lambda$, we perform  quantum simulations at four different $\lambda$ values, i.e.~800, 1100, 1300 and 1500 nm, respectively. These results are compared against those obtained with the classical approach, under the same set of laser parameters. 

\begin{figure}[h]
\centering
\includegraphics[width=0.7\textwidth]{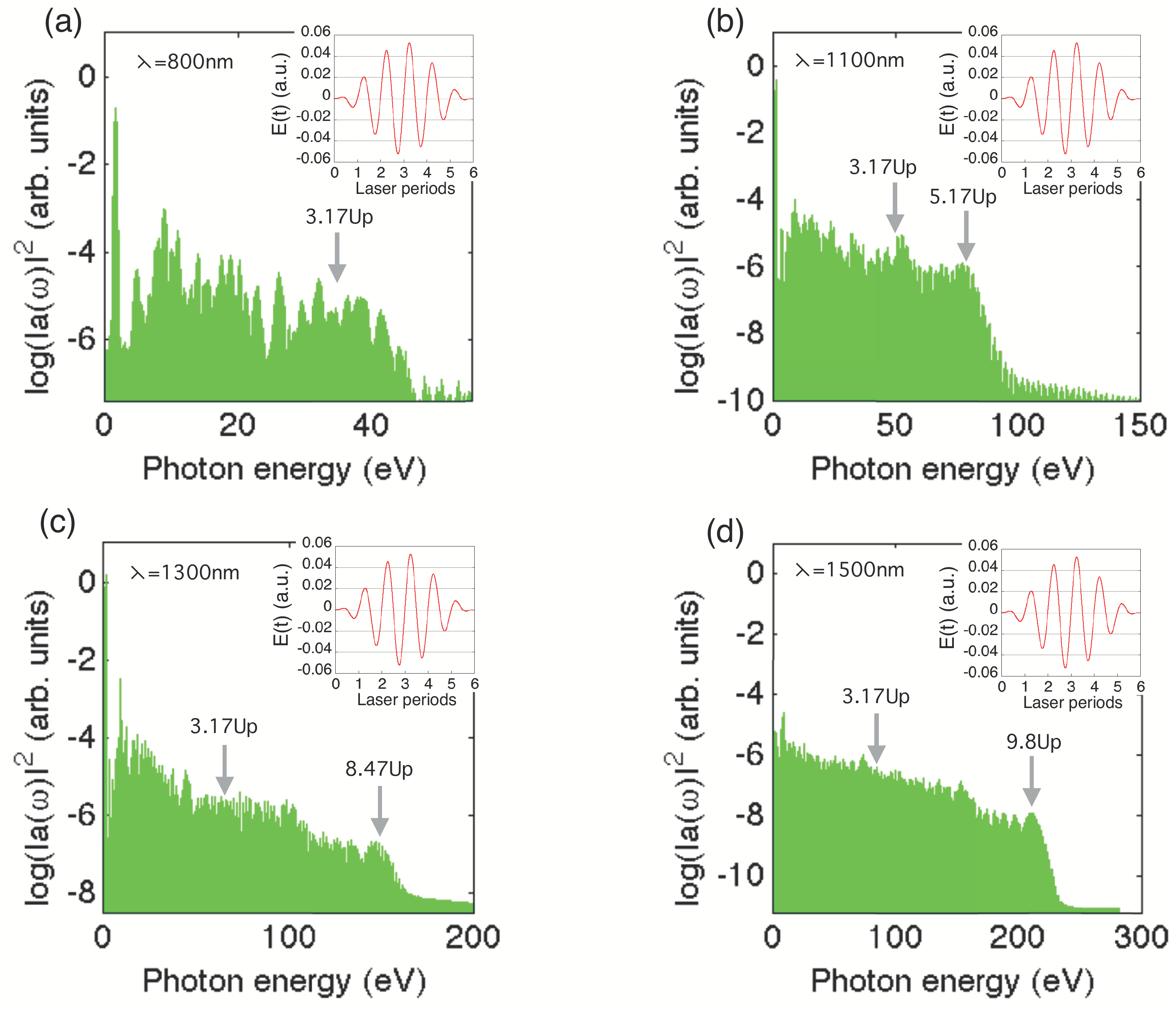}
\caption{1D-TDSE HHG spectra driven by the laser pulse described in Eq.~(\ref{eq:eq5}) at a plasmonic-enhanced laser intensity $I=1\times10^{14}$ W/cm$^{2}$. Four different wavelengths, namely (a) 800 nm, (b) 1100 nm, (c) 1300 nm and (d) 1500 nm are shown. The arrows in the panels indicate the conventional HHG cutoff ($3.17 U_p+I_p$) and the one obtained when a plasmonic-enhanced spatially inhomogeneous near-field is employed (note that the ionization potential $I_P$ of the argon atom, 15.76 eV,  is implicitly considered).}
\label{fig4} 
\end{figure}

\begin{figure}[h]
\centering
\includegraphics[width=0.4\textwidth]{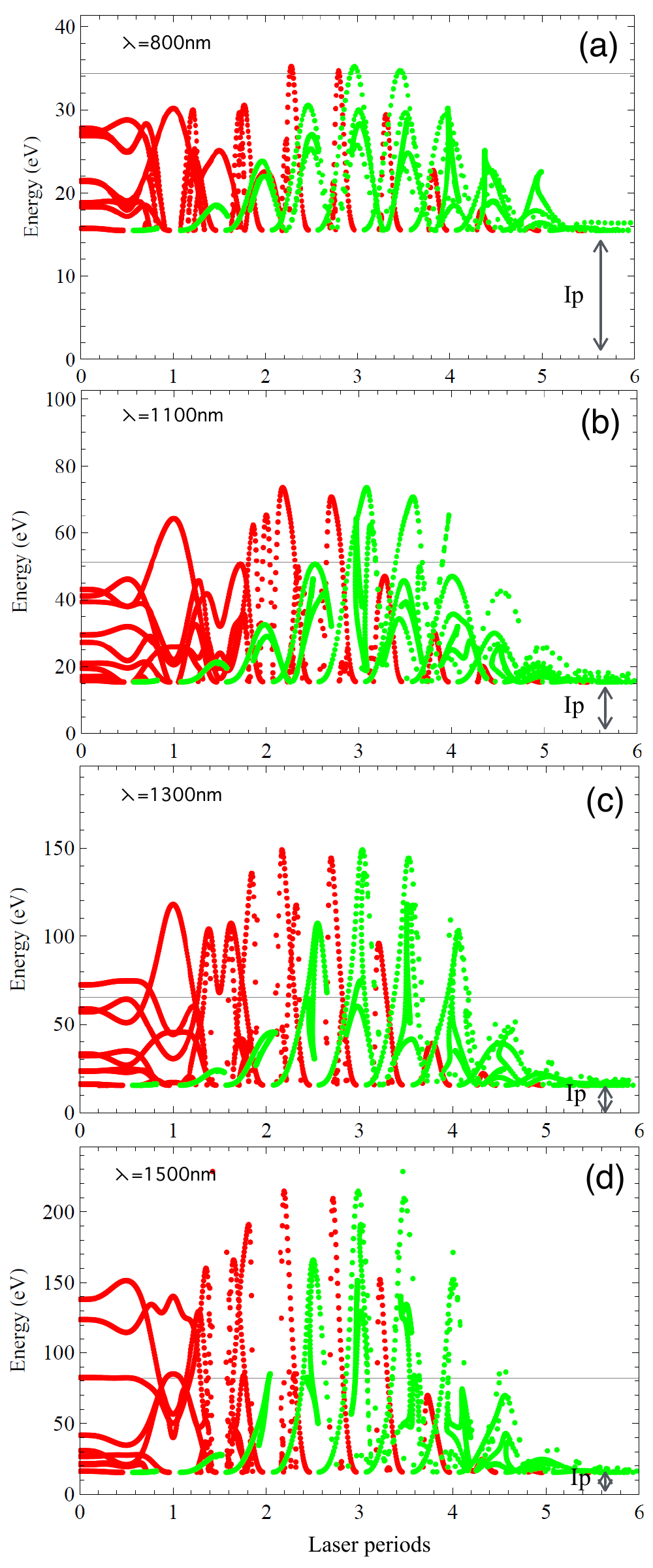}
\caption{Classical electron kinetic energy at recombination time as a function of the ionization (red) and recombination (green) times (for more details see e.g~\cite{cpc}). The system is driven by a spatial inhomogeneous field as showed in Fig.~\ref{fig2} and with the same parameters of Fig.~\ref{fig4}. The model atom is located at point A. The $I_p$ of argon (15.76 eV) and the classical HHG cutoff energy (grey solid line) are shown.}
\label{fig5} 
\end{figure}

In Fig.~\ref{fig4} we can observe the HHG spectra obtained by solving the 1D-TDSE for the four above-mentioned cases.  As it can be seen, for larger wavelengths a much higher photon energy (cutoff) is reached. This is not surprising, considering the HHG cutoff scales as $\lambda^2$ for spatial homogeneous fields. If we explore, however, the HHG cutoff dependence with the laser wavelength, it could be seen that, as the $\lambda$ increases, the cutoff behaves markedly different for spatial homogeneous and inhomogeneous fields. For $\lambda=800$ nm (Fig.~\ref{fig4}(a)) the HHG cutoff hardly differs if we use a spatial homogeneous or inhomogeneous field to drive the system. On the contrary, we find that for $\lambda =1100$, 1300 and 1500 nm the HHG cutoff energy shows an increment reaching values up to $5.17U_p$, $8.47U_p$ and $9.8U_p$, respectively (see Figs.~\ref{fig4}(b),~\ref{fig4}(c) and~\ref{fig4}(d), respectively). It is noticeable that the maximum cutoff energy increases faster than in the case of spatial homogeneous fields, i.e.~at a power larger than 2 with respect to the laser wavelength. On the other hand, considering the integration of the Newton-Lorentz equation, we can obtain the maximum energy for the electron at the recombination time with the parent ion.  These results are displayed in Fig.~\ref{fig5}. As it can be seen, the maximum values of velocity (kinetic energy) reached by the electrons follow the same behaviour observed in the quantum simulations. This once again supports the particular dependence of the cutoff energy with respect to the excitation laser wavelength.

\begin{figure}[h]
\centering
\includegraphics[width=0.7\textwidth]{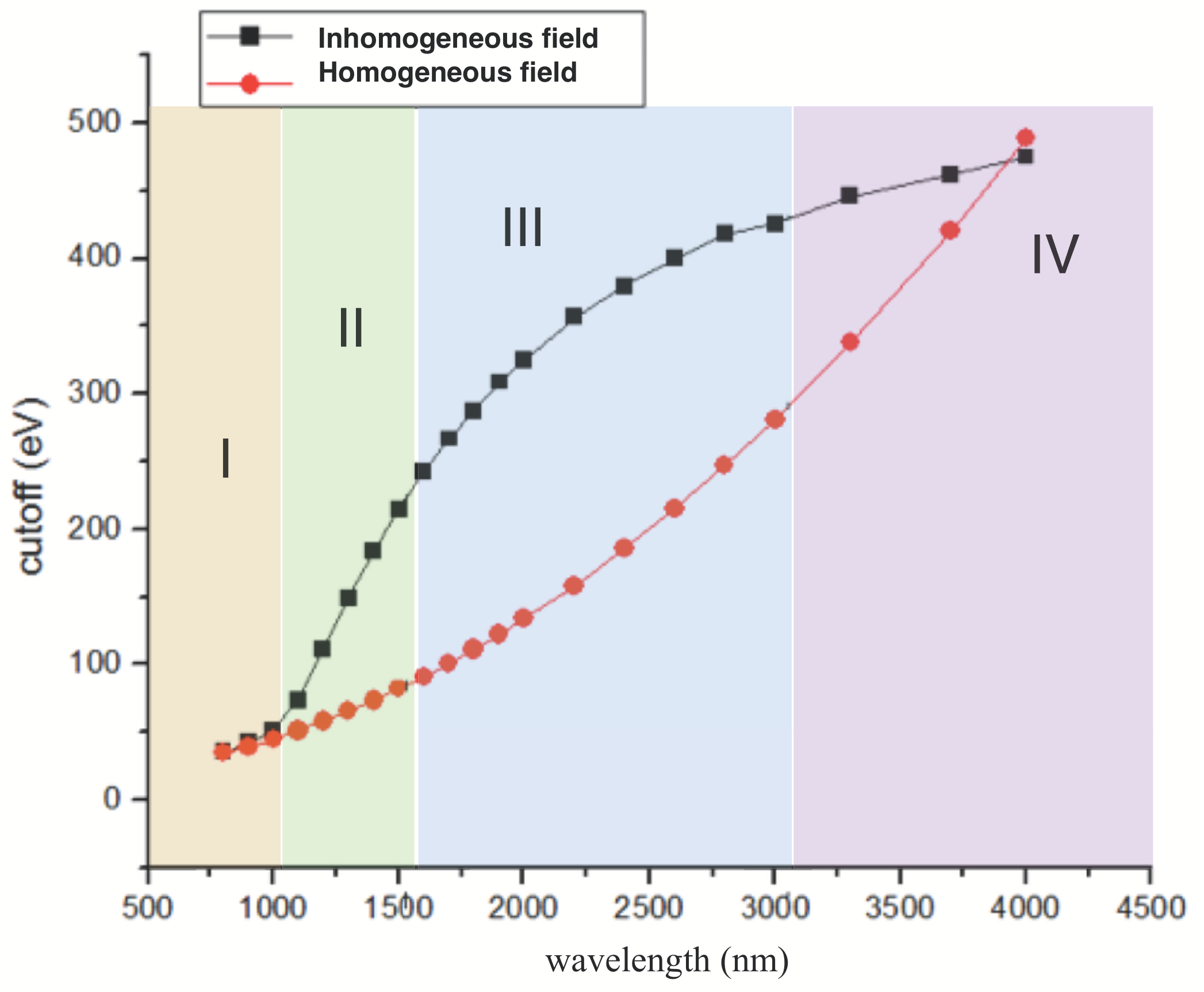}
\caption{Dependence of the HHG cutoff with the driving laser wavelength $\lambda$ for both spatial homogeneous (red circles) and inhomogeneous (black dots) fields. The Regions I-IV highlight the different functional dependences that describe the HHG cutoff behaviour (see the text for more details).}
\label{fig6} 
\end{figure}

In order to get more details about the above distinct behavior, we present in Fig.~\ref{fig6} the dependence of the HHG cutoff, obtained by means of the classical analysis, as a function of the excitation laser wavelength $\lambda$. We can thus distinguish four different regions, I to IV. In Region I (up to 1100 nm of excitation laser wavelength) the cutoff scales in a similar way as the homogeneous case. This is so because at these laser wavelengths the excursion of the laser-ionized electron is small (as a reference the red line corresponds to the case of a spatially homogeneous laser electric field) and, as a consequence, the spatial variation of the plasmonic-enhanced field is hardly  `probed' by the electron. In Region II, from 1100 to 1600 nm, the cutoff dependence scales with $\lambda$ at a power above 3; in this region the difference between spatial homogeneous and inhomogeneous fields is much more pronounced.

In Region III, above 1600 nm and up to 3000 nm, we observe a rise in the HHG cutoff lower than the classic law. This fact can be understood considering the limited ionization-acceleration region, as a consequence of the bound electric near-field created by the nanostructure. At these larger wavelengths the electron excursion is much more extended, so the electron reaches a spatial region where the electric field amplitude is strongly reduced according to the Gaussian-shaped profile fitting (see Fig.~\ref{fig2}). 

\begin{figure}[h]
\centering
\includegraphics[width=0.7\textwidth]{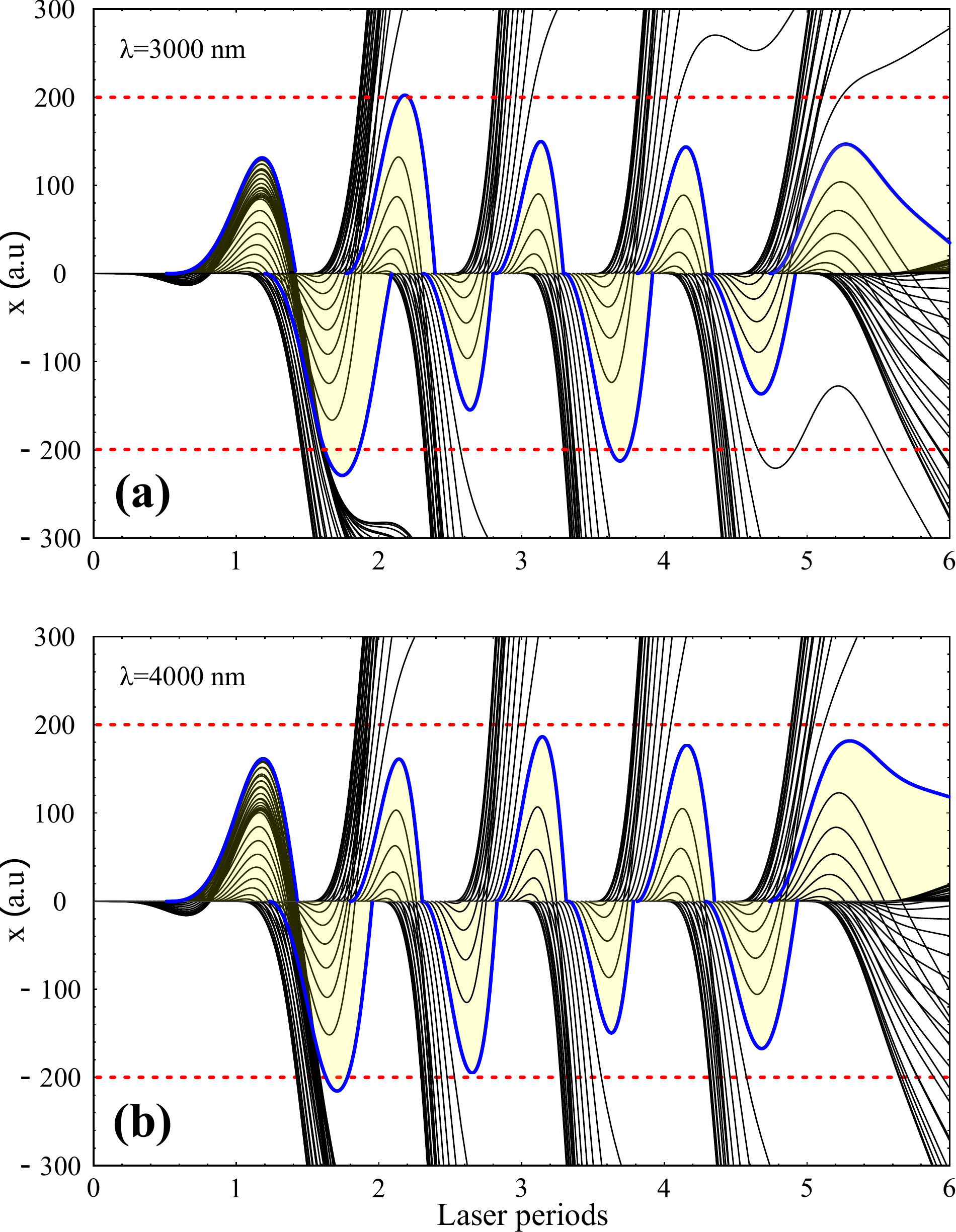}
\caption{Classical electron trajectories driven by a spatial inhomogeneous field at $\lambda=3000$ nm (a) and  $\lambda=4000$ nm (b), respectively. It can be observed that the span of trajectories that recombine (shaded region) never overcome $\pm \sim 200$ a.u. (red dashed line).}
\label{fig7} 
\end{figure}

Finally, in Region IV, defined for laser wavelengths larger than 3000 nm, a sort of saturation in the HHG cutoff energy is observed (compare with the spatial homogeneous electric field case, that follows the $\lambda^2$-law, irrespective of the laser wavelength range). To clarify this point, in Fig.~\ref{fig7} we show electron trajectories obtained by means of the classical approach.
In Fig.~\ref{fig7}(a) we present classical trajectories for a spatial inhomogeneous field at $\lambda=3000$ nm, meanwhile in Fig.~\ref{fig7}(b), $\lambda=4000$ nm. Interestingly, and for both spatial inhomogeneous cases, only few electron trajectories end at the parent ion, i.e.~the electron recombination occurs. This means that the efficiency of HHG driven by these long wavelength spatially inhomogeneous near-fields is relatively low.  Likewise, when a spatially inhomogeneous field drives the HHG process, two important facts can be extracted from Fig.~\ref{fig7}. On the one side, it is observed that for many electron trajectories there is no recombination. In turn, this fact results in two main physical effects: (i) a low HHG yield and (ii) a large amount of direct electrons in the above-threshold ionization (ATI) process. 

Both characteristics are associated with the spatial profile of the plasmonic-enhanced laser electric near-field. On the other side, it can be seen from Fig.~\ref{fig7} that, independently of the laser wavelength, the electron trajectories that recombine never overcome position values of $\pm$ 200 a.u., for this particular electric field amplitude (as it is indicated by black lines in Fig.~\ref{fig2}). This fact can be confirmed by comparing the electron trajectories for $\lambda=3000$ and 4000 nm, as it is shown in Fig.~\ref{fig7}(a) and Fig.~\ref{fig7}(b), respectively.  Additionally, this behavior can be understood taking into account the spatial extension of the electron trajectories developed at larger wavelengths. For these cases, the electron `feels' an electric field of low amplitude (see Fig.~\ref{fig2}) and, as a consequence, a limit in its kinetic energy appears. In summary, in spatially homogeneous fields the electron trajectories do not present limits and, as a consequence, they can reach larger displacements as larger wavelengths are used to drive the electron from the ionization point. In contrast, the trajectories for an electron driven by the studied Gaussian-shaped spatially inhomogeneous field, are constrained up to $\pm \sim 200$ a.u. This is because the plasmonic-enhanced electric near-field generated by the bow-tie shaped nanostructure is spatially bounded.

Additionally, the HHG cutoff energy reached at larger wavelengths, as we explained above, is limited due to the bound electric field, so the kinetic energy gained by the electron cannot overcome an asymptotic value close to $\sim500$ eV (Fig.~\ref{fig6}).

\section{Conclusions and outlook}

In conclusion, in this contribution we reported original features of the HHG driven by a plasmonic-enhanced near-field. On the one side, we observed that the HHG cutoff energy scales with $\lambda$ at a power above 3 in a noticeable range of laser wavelengths.  On the other side, in spatial inhomogeneous fields the electron classical trajectories showed several relevant characteristics: 
1) the electron displacements are limited by the extension of the bound plasmonic-enhanced spatially inhomogeneous near-field, independently of the excitation wavelength;
2) it is shown that the HHG cutoff energy scales at a power above 3. This fact is directly related to the electric field dependence with the position where the ionization-recombination process takes place and 3) we showed and confirmed, based on classical arguments, that most of the electron trajectories never recombine, so this process would indeed improve the direct-ATI generation.
 It is worth to be mentioned that the cutoff behaviour described in this work strongly depends on the functional form of the plasmonic-enhanced spatially inhomogeneous near-field proposed. As a consequence, it is expectable a different behavior if plasmonic-enhanced near-fields with other spatial properties are used to drive the HHG process.
Summarizing, the results presented in this paper could open new avenues for nanostructure engineering and the exploration of alternative approaches for plasmonic-enhanced HHG.

\section*{Acknowledgements}

This work was supported by the project ELI-Extreme Light Infrastructure-phase 2 (CZ.02.1.01/0.0/0.0/15\_008/0000162) from European Regional Development Fund. J. A. P.-H. and L. R. acknowledge to the Spanish Ministerio de Econom\'{\i}a y Competitividad (PALMA project FIS2016-81056-R), Laserlab-Europe (EU-H2020 654148) and Junta de Castilla y Le\'on project
CLP087U16. M. L. acknowledges MINECO (National Plan Grant: FISICATEAMO No. FIS2016-79508-P, SEVERO OCHOA No. SEV-2015-0522, FOQUS No. FIS2013-46768-P), Fundaci\'o Cellex
Generalitat de Catalunya (AGAUR Grant No. 2014
SGR874 and CERCA/Program), ERC AdG OSYRIS, EU
FETPRO QUIC, EU STREP EquaM (FP7/2007-2013 Grant
No. 323714) and the National Science Centre, Poland-Symfonia Grant No. 2016/20/W/ST4/00314. We also
acknowledge support from Spain's
MINECO (Grant No. DYNAMOLS FIS2013-41716-P).


\section*{References}

\begin{thebibliography}{99}


\bibitem{lhuiller} Ferray M, L'Huillier A, Li X F, Lompre L A, Mainfray G and Manus C 1988 Multiple-harmonic conversion of 1064 nm radiation in rare gases \textit{J. Phys. B: At., Mol. Opt. Phys.} \textbf{21} L31

\bibitem{protopapas} Protopapas M, Keitel C H and Knight P L 1997 Atomic physics with super-high intensity lasers \textit{Rep. Prog. Phys.} \textbf{60} 389

\bibitem{brabec} Brabec T and Krausz F 1997 Nonlinear optical pulse propagation in the single-cycle regime \textit{Phys. Rev. Lett.} \textbf{78} 3282

\bibitem{krausz} Krausz F and Ivanov M 2009 Attosecond physics \textit{Rev. Mod. Phys.} \textbf{81} 163

\bibitem{corku93A} Corkum P B 1993 Plasma perspective on strong field multiphoton ionization \textit{Phys. Rev. Lett.} \textbf{71} 1994

\bibitem{lewenstein94A} Lewenstein M, Balcou P, Ivanov M Yu, L��'Huillier A and Corkum P B 1994 Theory of high-harmonic generation by low-frequency laser fields \textit{Phys. Rev. A} \textbf{49} 2117

\bibitem{schaf93A} Schafer K J, Yang B, DiMauro L F and Kulander K C 1993 Above threshold ionization beyond the high harmonic cutoff \textit{Phys. Rev. Lett.} \textbf{70} 1599

\bibitem{spielmann} Spielmann C, Burnett N H, Sartania S, Koppitsch R, Schn\"urer M, Kan C, Lenzner M, Wobrauschek P and Krausz F 1997 Generation of coherent X-rays in the water window using 5-femtosecond laser pulses \textit{Science} \textbf{278} 661-664

\bibitem{tenio} Popmintchev T, et al. 2009 Phase matching of high harmonic generation in the soft and hard X-ray regions of the spectrum \textit{Proc. Nat. Acad. Sci.} \textbf{106} 10516-10521

\bibitem{tate_07A} Tate J, Auguste T, Muller H G, Salieres, P, Agostini P and DiMauro L F 2007 Scaling of wave-packet dynamics in an intense midinfrared field. \textit{Phys. Rev. Lett.} \textbf{98} 013901

\bibitem{frolov_08} Frolov M V, Manakov N L and Starace A F 2008 Wavelength scaling of high-harmonic yield: threshold phenomena and bound state symmetry dependence \textit{Phys. Rev. Lett.} \textbf{100} 173001

\bibitem{perez-hernandez09A} P\'erez-Hern\'andez J A, Roso L and Plaja L 2009 Harmonic generation beyond the Strong-Field Approximation: the physics behind the short-wave-infrared scaling laws. \textit{Opt. Exp.} \textbf{17} 9891-9903

\bibitem{strelkov} Strelkov V V, Sterjantov A F, Shubin N Y and Platonenko V T 2006 XUV generation with several-cycle laser pulse in barrier-suppression regime. \textit{J. Phys. B: At., Mol. Opt. Phys.} \textbf{39} 577

\bibitem{jose_oe} P\'erez-Hern\'andez J A, Roso L, Za\"ir A and Plaja L 2011 Valley in the efficiency of the high-order harmonic yield at ultra-high laser intensities. \textit{Opt. Exp.} \textbf{19} 19430

\bibitem{moreno1995} Moreno P, Plaja L, Malyshev V and Roso L 1995 Influence of barrier suppression in high-order harmonic generation. \textit{Phys. Rev. A} \textbf{51} 4746-4753 

\bibitem{farkas92} Farkas G and T\'oth C 1992 Proposal for attosecond light pulse generation using laser induced multiple-harmonic conversion processes in rare gases. \textit{Phys. Lett. A} \textbf{168} 447-450

\bibitem{hentschel01} Hentschel M, Kienberger R, Spielmann C, Reider G A, Milosevic N, Brabec T and Krausz F 2001 Attosecond metrology \textit{Nature} \textbf{414} 509

\bibitem{carrera} Carrera J J, Tong X M and Chu S I 2006 Creation and control of a single coherent attosecond XUV pulse by few-cycle intense laser pulses \textit{Phys. Rev. A} \textbf{74} 023404

\bibitem{chipperfield09} Chipperfield L E, Robinson J S, Tisch J W G and Marangos J P 2009 Ideal Waveform to Generate the Maximum Possible Electron Recollision Energy for Any Given Oscillation Period \textit{Phys. Rev. Lett.} {\bf 102} 063003

\bibitem{serrat} Serrat C and Biegert J 2010 All-regions tunable high harmonic enhancement by a periodic static electric field \textit{Phys. Rev. Lett.} \textbf{104} 073901

\bibitem{enriquelp} Neyra E, Videla F, P\'erez-Hern\'andez J A, Ciappina M F, Roso L and Torchia G A 2016 Extending the high-order harmonic generation cutoff by means of self-phase-modulated chirped pulses \textit{Las. Phys. Lett.} \textbf{13} 115303

\bibitem{enriqueepjd} Neyra E, Videla F, P\'erez-Hern\'andez J A, Ciappina M F, Roso L and Torchia G A 2016 High-order harmonic generation driven by chirped laser pulses induced by linear and non linear phenomena \textit{Eur. Phys. J. D} \textbf{70} 243

\bibitem{kim08} Kim S, Jin J, Kim Y-J, Park I-Y, Kim Y and Kim S-W 2008 High-harmonic generation by resonant plasmon field enhancement \textit{Nature} \textbf{453} 757

\bibitem{ropp} Ciappina M F, et al. 2017 Attosecond physics at the nanoscale 2017 \textit{Rep. Prog. Phys.} \textbf{80} 054401

\bibitem{sivise1} Sivis M, Duwe M, Abel B and Ropers C 2012 Nanostructure-enhanced atomic line emission \textit{Nature} \textbf{485} E1-E3

\bibitem{kime2} Kim S, Jin J, Kim Y-J, Park I-Y, Kim Y and Kim S-W 2012 Kim et al. reply \textit{Nature} \textbf{485} E1-E3

\bibitem{sivis2013}  Sivis M, Duwe M, Abel B and Ropers C 2013 Extreme-ultraviolet light generation in plasmonic nanostructures \textit{Nat. Phys.} \textbf{9} 304-309

\bibitem{han2016} Han S, Kim H, Kim Y W, Kim Y-J, Kim S, Park I-Y and Kim S-W 2016 High-harmonic generation by field enhanced femtosecond
pulses in metal-sapphire nanostructure \textit{Nat. Commun.} \textbf{7} 13105

\bibitem{prl} P\'erez-Hern\'andez  J A, Ciappina M F, Lewenstein M, Roso L and Za\"ir A 2013 Beyond carbon K-edge harmonic emission using a spatial and temporal synthesized laser field \textit{Phys. Rev. Lett.} \textbf{110} 053001

\bibitem{cpc}Ciappina M F, P\'erez-Hern\'andez J A and Lewenstein M 2014 ClassSTRONG: Classical simulations of strong field processes \textit{Comp. Phys. Comm.} \textbf{185} 398-406

\bibitem{husakou11} Husakou A, Im S J and Herrmann J 2011 Theory of plasmon-enhanced high-order harmonic generation in the vicinity of metal nanostructures in noble gases \textit{Phys. Rev. A} \textbf{83} 043839

\bibitem{yavuz12} Yavuz I, Bleda E A, Altun Z and Topcu T 2012 Generation of a broadband XUV continuum in high-order-harmonic generation by spatially inhomogeneous fields \textit{Phys. Rev. A} \textbf{85} 013416

\bibitem{marcelo12} Ciappina M F, Biegert J, Quidant R and Lewenstein M 2012 High-order-harmonic generation from inhomogeneous fields \textit{Phys. Rev. A} \textbf{85} 033828

\bibitem{marcelooe} Ciappina M F, Ac�imovic� S S, Shaaran T, Biegert J, Quidant R and Lewenstein M 2012 Enhancement of high harmonic generation by confining electron motion in plasmonic nanostrutures \textit{Opt. Exp.} \textbf{20} 26261-26274

\bibitem{Rsoft} FullWAVE-Rsoft User Guide 2008 RSoft Design Group, Inc. 400 Executive Blvd., Suite 100, Ossining, New York 10562.

\bibitem{yavuz2013} Yavuz I 2013 Gas population effects in harmonic emission by plasmonic fields \textit{Phys. Rev. A} \textbf{87} 053815

\bibitem{potential} Su Q and Eberly J H 1991 Model atom for multiphoton physics \textit{Phys. Rev. A} \textbf{44} 5997

\bibitem{tahir12} Shaaran T, Ciappina M F and Lewenstein M 2012 Quantum-orbit analysis of high-order-harmonic generation by resonant plasmon field enhancement \textit{Phys. Rev. A} \textbf{86} 023408





\end{thebibliography}
\end{document}